\documentclass[preprint,aps,shownopacs,preprintnumbers,amsmath,amssymb,nofootinbib,epsf,epsfig]
{revtex4}
\usepackage[latin2]{inputenc}
\usepackage{amsmath}
\usepackage{amsfonts}
\usepackage{amssymb}
\usepackage{graphicx}
\usepackage{slashed}
\begin{document}

\title{Effect of Non-commutativity of space-time on Thermodynamics of Photon gas}
\author{Ravikant Verma$^{a,b}$\footnote{Short Time Visit at University of Hyderabad}}
\email{ravikant.uohyd@gmail.com, ravikant.verma@bose.res.in}
\author{Partha Nandi$^c$}
\email{parthanandi@bose.res.in}

\affiliation{$^a$School of Physics,  University of Hyderabad, Hyderabad-500046,  India\\
$^b$Naraina Vidya Peeth Engineering and Management Institute, Panki, Kanpur-208020, India\\
$^c$S. N. Bose National Centre for Basic Science, JD Block, Sector III, Salt Lake, Kolkata-700106, India}

\begin{abstract}
Doubly special relativity(DSR) introduce a minimal length scale i.e. the Planck length scale, which is independent length scale in addition to the speed of light in the normal special theory of relativity(STR). Doubly special relativity leads to study the $\kappa$-Minkowski space-time. In this paper, we present the result of our investigation on the thermodynamics of photon gas in the $\kappa$-Minkowski space-time. For studying this, we start with the $\kappa$-deformed dispersion relation and keep terms upto first order in the deformation parameter $a$ and we study that how does $\kappa$-deformed dispersion relation affect the thermodynamics of photon gas. In the limit, deformation parameter $a \rightarrow 0$, we get back all the results in the special theory of relativity(STR)\cite{partition}.
\end{abstract}
\maketitle
\section{Introduction}
Theory of quantum gravity model suggests to existence of the minimal scale of length i.e. the Planck's length $\l_p =\sqrt{\frac{G\hbar}{c^3}}$ (or corresponding energy scale $E_p=\sqrt{\frac{\hbar c^5}{G}}$ which is maximum energy of the particle), but existence of the fundamental length scale i.e. Planck's length is contradictory to the special theory of relativity due to the Lorentz length contraction. Therefore Amelino-Camilia et al. suggest a way to handle this contradiction to the special theory of relativity and proposed a new theory which is called the doubly special relativity (DSR)\cite{d1,d2,d3,d4}. In this new proposed theory Planck's length is the fundamental length and observer independent in addition to the speed of light. Various aspects of physical models
incorporating DSR where analysed in\cite{d11,d12,d13}. Doubly special relativity leads to study the $\kappa$-deformed space-time\cite{fgrd1,fgrd2,fgrd3,fgrd4,fgrd5} which is one of the example of non-commutative space-time and the corresponding symmetric algebra is deformed which is called the $\kappa$-Poincare algebra.

Different theories on $\kappa$-deformed space-time have been constructed and studied in the last couple of years\cite{fgrd5,1a,1,2,3,4,5,6,7,8,9,10,10A}. Since dispersion relations are modified due to the $\kappa$-deformation of the underlying space-time\cite{fgrd5}, it is interesting to see how the thermodynamics of a photon gas affected by such a modification. The effect of various kinds of deformations of the dispersion relations on the thermodynamics of a photon gas has been studied before\cite{11,12,13,14,15,16}.

The motivation of this work is to carry out the thermodynamic properties of the photon gas in the $\kappa$-deformed space-time. Such a investigation has been carried out with a modified dispersion relation in the different scenario. Our present study would therefore help us to compare our results with those studied in\cite{11,12,13,14,15,16}. It should be noted, however, that we shall consider the $\kappa$-deformed dispersion relation in our analysis. This in turn would incorporate the effects of $\kappa$-deformed space-time on the partition function of photon gas. Here note that the quantum distribution functions, whether Fermi-Dirac distribution or Bose-Einstein distribution reduce to the Maxwell-Boltzmann distribution in the high temperature. In this paper, we work in the high temperature limit.

This paper is organised as follows. In next section, we give the brief summary of the $\kappa$-deformed space-time\cite{fgrd1,fgrd2,fgrd3,fgrd4,fgrd5} and write the $\kappa$-deformed dispersion relation upto first order in the deformation parameter. By using the $\kappa$-deformed dispersion relation upto first order in the deformation parameter, we drive important expression for the partition function of photon gas in section 3. In the section 4, we go on to study the thermodynamic properties of the photon gas in the $\kappa$-deformed space-time by using $\kappa$-deformed partition function. We calculate the expression for free energy, pressure, equation of state, entropy, internal energy, energy density and specific heat of the photon gas in $\kappa$-deformed space-time and we make comparisons between our results, Magueijo-Smolin(MS) model results\cite{14} and results in special theory of relativity\cite{partition} by plotting the graphs between the thermodynamics variables versus temperature. In section 5, we discuss about the MS model and our considered model. We make the final remark in the section 6.

\section{$\kappa$-Minkowski space-time}
In this section, we give the brief summary of $\kappa$-deformed space-time\cite{fgrd1,fgrd2,fgrd3,fgrd4,fgrd5}, which is an example of a non-commutative space-time whose coordinates obey commutation relations,
\begin{equation}
 [\hat{x}^i, \hat{x}^j]=0,~~ [\hat{x}^0, \hat{x}^i]=i a \hat{x}^i.\label{kappacom}
\end{equation}
Here, $a$ is the deformation parameter and has the dimension of length and it is order of Planck length scale(in the limit $a\rightarrow 0$, we get back the above deformed commutation relation in the commutative space-time). The above algebra can be written in a covariant form in terms of Minkowski metric $\eta_{{\mu}{\nu}}$ = diag(-1,1,1,...,1) as
\begin{equation}
[\hat{x}^{\mu}, \hat{x}^{\nu}]= iC^{\mu\nu\lambda}\hat{x}_{\lambda}=i(a^{\mu}\eta^{\nu\lambda}-a^{\nu}\eta^{\mu\lambda})\hat{x}_{\lambda}
\end{equation} 
where $a_{\mu} (\mu,\nu = 0,1,....,n-1)$ are real and dimensionful constants with $a_0 = a$ and $a_i=0$ for $i=1,2,....n-1$. The $\kappa$-Minkowski space-time coordinates $\hat{x}_\mu$ are defined in terms the commutative coordinates $x_\mu$ and corresponding derivatives $\partial_\mu$ as $\hat{x}_\mu=x^\alpha \phi_{{\alpha}{\mu}}(\partial)$.
 In this realization, the explicit form of $\hat{x}_i$ and $\hat{x}_0$ are given as
\begin{eqnarray}
\hat{x}_i=x_i \varphi(A),~~ \hat{x}_0=x_0 \psi(A) + i a x_i \partial_i \gamma(A),\label{partial}
\end{eqnarray}
where $A=- ia \partial_0$. Using Eqn.(\ref{partial}) in Eqn.(\ref{kappacom}), we obtain
\begin{equation}
\frac{\varphi^\prime}{\varphi}\psi=\gamma(A)-1
\end{equation}
where $\varphi^\prime=\frac{d\varphi}{dA}$ satisfying the boundary conditions $\varphi(0)=1, \psi(0)=1, \gamma(0)=\varphi^\prime (0)+1$ is finite and $\gamma,\varphi,\psi $ are positive functions.

 These coordinates satisfy the condition that in particular $[\partial_\mu , \hat{x}_\nu]=\phi_{{\mu}{\nu}}(\partial)$, we find 
\begin{equation}
[\partial_i , \hat{x}_j]=\delta_{{i}{j}} \varphi(A) ,~~[\partial_i , \hat{x}_0]=i a \partial_{i} \gamma(A) ,~~ [\partial_0 , \hat{x}_i]=0,
\end{equation}
and
\begin{equation}
[\partial_0 ,\hat{x}_0]=\eta_{00}=-1.
\end{equation}
Let $M_{\mu\nu}$ denote the rotation and boost generators of the $\kappa$-Poincare algebra. We require that the commutation relations between $M_{\mu\nu}$ should be same as the usual Poincare algebra and also that between $M_{\mu\nu}$ and the $\kappa$-space-time coordinates $\hat{x}_{\mu}$ should be linear functions of $\hat{x}_{\mu}$,~$M_{\mu\nu}$ and deformation parameter $a_\mu$. We also demand that these generators has the correct commutative limit, lead to just two class of possible realizations. They are parameterized by $\psi =1$ and $\psi = 1 + 2A$. Thus one can get\cite{fgrd5}
\begin{equation}
[M_{\mu\nu},\hat{x}_{\lambda}]=\hat{x}_{\mu}\eta_{\nu\lambda}-\hat{x}_{\nu}\eta_{\mu\lambda}-i(a_\mu M_{\nu\lambda}-a_\nu M_{\mu\lambda}).
\end{equation}
\begin{equation}
[M_{i0}, \hat{x}_0] = -\hat{x}_i +ia M_{i0}
\end{equation}
\begin{equation}
[M_{i0}, \hat{x}_j] = -\delta_{ij}\hat{x}_0 +ia M_{ij}
\end{equation}
The symmetry algebra of the underlying $\kappa$-space-time generated by $M_{\mu \nu}$ and $D_\mu$ known as the undeformed $\kappa$-Poincare algebra. Their generators $D_{\mu}$ and $M_{{\mu}{\nu}}$ obey\cite{fgrd1,fgrd2,fgrd3,fgrd4,fgrd5}
\begin{equation}
[M_{{\mu}{\nu}} , D_{\lambda}]=\eta_{{\nu}{\lambda}} D_\mu - \eta_{{\mu}{\lambda}} D_\nu ,~~ [D_\mu , D_\nu]=0,\label{dirac4}
\end{equation}
\begin{equation}
[M_{{\mu}{\nu}}, M_{{\lambda}{\rho}}]=\eta_{{\mu}{\rho}} M_{{\nu}{\lambda}} + \eta_{{\nu}{\lambda}} M_{{\mu}{\rho}} - \eta_{{\nu}{\rho}} M_{{\mu}{\lambda}} - \eta_{{\mu}{\lambda}} M_{{\nu}{\rho}}.\label{dirac5}
\end{equation}
In the above, we use $\eta_{\mu \nu}$ = diag$(-1, 1, 1, 1)$. Here note that the (Dirac) derivatives $D_\mu$ above, transform as vectors (unlike the usual derivative operators in the $\kappa$-Minkowski space-time). But here the realization of the generators do have $a$ dependent terms.

The explicit form of generators(for the realization $\psi=1$) are given by\cite{fgrd5}
\begin{equation}
M_{ij}=x_{i}\partial_{j}-x_{j}\partial_{i},\label{dg1}
\end{equation}
\begin{equation}
M_{i0}=x_i \partial_0 \varphi \frac{e^{2A}-1}{2A}-x_0 \partial_i \frac{1}{\varphi}+iax_i \bigtriangledown^2 \frac{1}{2\varphi}-iax_k \partial_k \partial_i \frac{\gamma}{\varphi},\label{dg2}
\end{equation}
and
\begin{equation}
D_i=\partial_i \frac{e^{-A}}{\varphi},~~ D_0=\partial_0 \frac{ \sinh A}{A} + i a \bigtriangledown^2 \frac{e^{-A}}{2 \varphi^2}. \label{dr}
\end{equation}
Where  $\bigtriangledown^2=\partial_k \partial_k$ and here $i,j,k=1,2,...n-1$. The undeformed $\kappa$-Poincare algebra in Eqns.(\ref{dirac4},\ref{dirac5}) have same form as the Poincare algebra in commutative space-time but form of generators get modified due to the $\kappa$-deformation of space-time. The explicit form of these generators are given in Eqn.(\ref{dg1}), Eqn.(\ref{dg2}) and Eqn.(\ref{dr}).

The Casimir of undeformed $\kappa$-Poincare algebra, $D_{\mu}D^{\mu}$, can be expressed as 
\begin{equation}
D_{\mu}D^{\mu}=\square (1 + \frac{a^2}{4} \square).\label{casmir}
\end{equation}
The $\square$ operator in the above satisfy
\begin{equation}
[M_{{\mu}{\nu}} , \square]=0, [\square,\hat{x}_\mu]=2D_\mu ,
\end{equation}
and the explicit form of the $\square$ operator is
\begin{equation}
\square=\bigtriangledown^2 \frac{e^{-A}}{2 \varphi^2} + 2 \partial_0^2 \frac{(1-\cosh A)}{A^2}.\label{box}
\end{equation}
Note that the Klein-Gordon equation in the commutative space-time can be expressed as $p_\mu p^\mu \Phi(x)=0$, where $p_\mu p^\mu$ is the Casimir operator of the Poincare algebra. We generalise this to $\kappa$-space-time and obtain the $\kappa$-deformed Klein-Gordon equation. Thus the generalised Klein-Gordon equation, invariant under undeformed $\kappa$-Poincare algebra, is written, using the Casimir of the undeformed $\kappa$-Poincare algebra as\cite{fgrd5} 
\begin{equation}
\square (1 + \frac{a^2}{4} \square) \Phi(x) - m^2 \Phi(x)=0. \label{kg}
\end{equation}
The deformed dispersion relation resulting from Eqn.(\ref{kg}) is
\begin{equation}
\frac{4}{a^2} \sinh^2\left(\frac{ap_0}{2}\right) - p_{i}^{2} \frac{e^{-ap_0}}{\varphi^2(ap_0)} + \frac{a^2}{4} \left[\frac{4}{a^2} \sinh^2\left(\frac{ap_0}{2}\right) - p_{i}^{2} \frac{e^{-ap_0}}{\varphi^2(ap_0)}\right]^2 = m^2.\label{D}
\end{equation}
In the limit $a\rightarrow 0$, we get back the dispersion relation in commutative space-time as $E^2 - p_i^2 =m^2$. For the choice of $\varphi=e^{-ap_0}$\cite{fgrd5,8} above $\kappa$-deformed dispersion relation take the form upto first order in the deformation parameter $a$ as
\begin{equation}
E^2 - p^2 (1+aE) + {\cal{O}}(a^2)=m^2
\end{equation}

\section{$\kappa$-deformed Partition function of Photon gas}
In this section, we present one of main result of the paper and see that how does $\kappa$-deformation of the space-time affect the partition function of photon gas which is the crucial result to study the thermodynamic properties of photon gas in the $\kappa$-deformed space-time. For studying this, we start with the $\kappa$-deformed dispersion relation of photon gas upto first order in the deformation parameter $a$ which is given as
\begin{equation}
E^2 = p^2(1+aE)
\end{equation}
\begin{equation}
E = p(1+aE)^\frac{1}{2}
\end{equation}
\begin{equation}
E = p\Big(1+\frac{aE}{2}\Big).
\end{equation}
Since $l_p=\frac{h c}{2\pi E_p}$ and now here note that deformation parameter $a$ has the dimension of length and is the order of Planck length scale and if we take $E_{p}=\lambda$ as a upper bound of energy scale i.e Planck energy scale then $a=\frac{h c}{2\pi \lambda}$ or $a=\frac{1}{2\pi \lambda}$ if we consider $h=1=c$. We can re-write above equation as
\begin{equation}
p=\frac{E}{1+\frac{E}{4\pi\lambda}}.
\end{equation}
or
\begin{equation}
p_{max}=\frac{E_{p}}{1+\frac{E_{p}}{4\pi\lambda}} = \frac{\lambda}{1+\frac{\lambda}{4\pi\lambda}}=\frac{\lambda}{1+\frac{7}{88}} =\frac{88}{95}\lambda .
\end{equation}
In the model which we considered, momentum in the upper bound of energy scale is given by $p_{max}=\frac{88}{95}\lambda$. Here we note that in the high temperature, the quantum distribution functions reduce to the classical distribution function and in this paper, we work with high temperature limit. It is very crucial to get an expression for the partition function to studying the thermodynamic properties of the particles. The single particle partition function $Z_1(T,V)$ is defined as\cite{partition} 
\begin{equation}
Z_1(T,V) = 4\pi V \int_0^\infty p^2 e^{-\beta E} dp,
\end{equation}
where $\beta =\frac{1}{K_B T}$, $K_B$ is the Boltzmann constant and T is the temperature of the particle. In the $\kappa$-Minkowski space-time, the single particle partition function $\tilde{Z}_1(T,V)$ is defined as
\begin{equation}
\tilde{Z}_1(T,V) = 4\pi V \int_0^{\frac{88}{95}\lambda} p^2 e^{-\beta E} dp.
\end{equation}
Note that $\lambda$ is the upper bound of the energy scale in the $\kappa$-deformed space-time. In the limit $\lambda \rightarrow \infty$, we get back the commutative result i.e. the result in special theory of relativity(STR)\cite{partition}. Using the $\kappa$-deformed dispersion relation for photons $p=\frac{E}{1+\frac{E}{4\pi\lambda}}$ in the above equation, we change the integration from $p$ to $E$ and we get
\begin{equation}
\tilde{Z}_1(T,V) = 4\pi V \int_0^\lambda E^2e^{-\beta E} \Big(1+\frac{E}{4\pi\lambda}\Big)^{-4} dE
\end{equation}
Since $\lambda$ is the upper bound of energy and it is very large therefore we can make an approximation and re-write above equation as
\begin{equation}
\tilde{Z}_1(T,V) = 4\pi V \int_0^\lambda E^2 \Big(1-\frac{E}{\pi\lambda}\Big)e^{-\beta E} dE,
\end{equation}
again above equation, we can re-write as
\begin{equation}
\tilde{Z}_1(T,V) = 4\pi V \Bigg[\frac{\partial^2}{\partial\beta^2} \int_0^\lambda e^{-\beta E}dE +\frac{1}{\pi\lambda}\frac{\partial^3}{\partial\beta^3} \int_0^\lambda e^{-\beta E}dE \Bigg],\label{int2}
\end{equation}
\begin{equation}
\int_0^\lambda e^{-\beta E}dE = \frac{1- e^{-\beta \lambda}}{\beta}\label{int1}
\end{equation}
Using eqn.(\ref{int1}) in eqn.(\ref{int2}) and differentiate with respect to $\beta$, we find the single particle $\kappa$-deformed partition function as
\begin{equation}
\tilde{Z}_1(T,V) = 4\pi V \Bigg[\frac{2}{\beta^3} - \frac{6}{\pi\lambda\beta^4} - \frac{e^{-\beta\lambda}}{\beta^4}\Big( \frac{2\beta}{7\pi} - \frac{6}{\pi\lambda} + \frac{23}{7\pi}\lambda\beta^2 +\frac{15}{7\pi} \lambda^2 \beta^3 \Big)        \Bigg] +{\cal{O}}\bigg(\frac{1}{\lambda^2}\bigg),\label{fr}
\end{equation}
In the limit $\lambda\rightarrow \infty$, we get back the commutative result $Z_1 = \frac{8\pi V}{\beta^3}$.

The total partition function for the $N$-particles system in the Maxwell-Boltzmann statistics will be\cite{gorji}
\begin{equation}
\tilde{Z}_N(T,V) = \frac{1}{N!} \Big[ \tilde{Z}_1(T,V)\Big]^N 
\end{equation}
Since here we have considered the photon as a massless relativistic classical particle. We are not considering quantum nature of photon\cite{gorji} therefore we are using canonical ensemble for fixed $N$ number of particles.

Thus, we have obtained the expression for the partition function for the photon gas in the $\kappa$-deformed space-time. Now we go on to study various thermodynamic properties of photon gas in a theory where an observer independent fundamental energy scale is present.

\section{Thermodynamics of $\kappa$-deformed Photon gas }
In the previous section, we have obtained the $\kappa$-deformed partition function for the photon gas but in the present section, we will study the certain aspects of thermodynamics of photon gas by using the $\kappa$-deformed partition function. The $\kappa$-deformed free energy of photon gas is given by
\begin{equation}
\tilde{F}= -K_B T \ln [\tilde{Z}_N(T,V)]=-K_B T\ln\Bigg[\frac{\tilde{Z}_1^N(T,V)}{N!}\Bigg]
\end{equation}
Now we use the Stirling's formula
\begin{equation}
\ln [N!]\approx N\ln[N] -N\nonumber
\end{equation}
in the above equation and by using the $\kappa$-deformed particle partition function for photon gas (see eqn.(\ref{fr})), we obtained
\begin{equation}
\tilde{F}=-NK_B T \Bigg[1+ \ln\Bigg( \frac{4\pi V}{N}{\cal{G}}\Bigg)\Bigg] +{\cal{O}}\bigg(\frac{1}{\lambda^2}\bigg)
\end{equation}
where 
\begin{eqnarray}
{\cal{G}}&=& 2K_B^3 T^3-\frac{21K_B^4 T^4}{11\lambda}-e^{-\frac{\lambda}{K_BT}}\Bigg( \frac{1}{11} K_B^3 T^3 -\frac{21K_B^4 T^4}{11\lambda}\nonumber\\ &+&\frac{23}{22}\lambda K_B^2 T^2 + \frac{15}{22}\lambda^2 K_B T \Bigg).
\end{eqnarray}
Here note that in the limit $\lambda \rightarrow \infty$, we get back the commutative result as\cite{partition}
\begin{equation}
F=-NK_B T \Bigg[1+ \ln\Bigg( \frac{8\pi V}{N}K_B^3T^3 \Bigg)\Bigg]. 
\end{equation}
Now we have the expression for the free energy and this can be used to calculate the pressure in the $\kappa$-deformed space-time as
\begin{equation}
\tilde{P}=-\bigg(\frac{\partial \tilde{F}}{\partial V} \bigg)_{T,N} = \frac{NK_B T}{V}
\end{equation}
or
\begin{equation}
PV = NK_B T
\end{equation}
Thus, the equation of state does not affected due to the $\kappa$-deformation of space-time.

As we have the expression for the free energy in the $\kappa$-deformed space-time and we can use it to find the effect of $\kappa$-deformed space-time on the entropy as
\begin{equation}
\tilde{S} = -\bigg(\frac{\partial \tilde{F}}{\partial T} \bigg)_{V,N}
\end{equation}
\begin{equation}
\tilde{S} = NK_B \Bigg[ 1+ \ln\bigg(\frac{4\pi V}{N}{\cal{G}}\bigg)\Bigg] + \frac{NK_B T}{{\cal{G}}}\frac{\partial {\cal{G}}}{\partial T}+{\cal{O}}\bigg(\frac{1}{\lambda^2}\bigg).
\end{equation}
Here,
\begin{eqnarray}
\frac{\partial {\cal{G}}}{\partial T} &=& 6K_B^3 T^2 - \frac{84K_B^4 T^3}{11\lambda} - e^{-\frac{\lambda}{K_B T}}\bigg( -\frac{18}{11} K_B^3 T^2 -\frac{84K_B^4 T^3}{11\lambda}\nonumber\\ &+& \frac{24}{11}\lambda K_B^2 T + \frac{19}{11} \lambda^2 K_B +\frac{15}{22} \frac{\lambda^3}{T}\bigg).
\end{eqnarray}
In the limit $\lambda \rightarrow \infty$, we get back the commutative result as in special theory of relativity as
\begin{equation}
S= NK_B \Bigg[ 4 + \ln \bigg( \frac{8\pi V}{N}(K_BT)^3\bigg)\Bigg].
\end{equation}
As internal energy is related to the free energy and entropy $U=F+TS$. Free energy and entropy of photon gas are modified due to the $\kappa$-deformation of space-time then we expect to modification in the internal energy due to the $\kappa$-deformation of the space-time as
\begin{equation}
\tilde{U} = \tilde{F} + T \tilde{S} = \frac{NK_B T^2}{{\cal{G}}}\frac{\partial {\cal{G}}}{\partial T}+{\cal{O}}\bigg(\frac{1}{\lambda^2}\bigg).\label{U}
\end{equation} 
In the limit $\lambda\rightarrow\infty$, we get back the result in the special theory of relativity as
\begin{equation}
U=3NK_BT.
\end{equation}
Energy density of the system $\rho$ is defined as
\begin{equation}
\rho = \frac{U}{V},
\end{equation}
since internal energy is modified in $\kappa$-deformed space-time then we also expect that the modification in the expression for the energy density. The modification between energy density and pressure is obtained as
\begin{equation}
\tilde{P}=\frac{\tilde{\rho} {\cal{G}}}{T\frac{\partial {\cal{G}}}{\partial T}}+{\cal{O}}\bigg(\frac{1}{\lambda^2}\bigg),
\end{equation}
In the limit $\lambda\rightarrow\infty$, we get back the commutative result as
\begin{equation}
P=\frac{1}{3}\rho.
\end{equation}
The expression for the $\kappa$-deformed specific heat of the photon gas is written as
\begin{equation}
\tilde{C}_V = \bigg(\frac{\partial \tilde{U}}{\partial T}\bigg)_V = T \bigg(\frac{\partial\tilde{S}}{\partial T}\bigg)_V
\end{equation} 
\begin{equation}
\tilde{C}_V =\frac{2NK_BT}{{\cal{G}}}\frac{\partial {\cal{G}}}{\partial T}+\frac{NK_BT^2}{{\cal{G}}}\frac{\partial^2 {\cal{G}}}{\partial T^2} - \frac{NK_BT^2}{{\cal{G}}^2}\bigg(\frac{\partial {\cal{G}}}{\partial T}\bigg)^2 + {\cal{O}}\bigg(\frac{1}{\lambda^2}\bigg),
\end{equation}
where,
\begin{eqnarray}
\frac{\partial^2 {\cal{G}}}{\partial T^2}&=&12K_B^3T -\frac{252K_B^4T^2}{11\lambda} - e^{-\frac{\lambda}{K_BT}}\Bigg( -\frac{120}{11} K_B^3T -\frac{252K_B^4T^2}{11\lambda}\nonumber\\ &+& \frac{6}{11}\lambda K_B^2 +\frac{23\lambda^2K_B}{11T}+\frac{23\lambda^3}{22T^2}+\frac{15\lambda^4}{22K_BT^3}\Bigg).
\end{eqnarray}
In the limit $\lambda\rightarrow\infty$, we get back the specific heat in the commutative space-time as $C_V=3NK_B$.

In the figure I-III, we plot the graph of entropy, internal energy and specific heat versus temperature for three cases: special theory of relativity\cite{partition}, Magueijo-Smolin(MS) model\cite{14} and the $\kappa$-deformed space-time (our considered model in this paper). Here we use Planck units and corresponding parameters have the following values: $K_B=1,~\lambda=10000,~N=10000$ and $V=0.01$. In figure-I, it is clear that at low temperature, entropy of photon gas is same for MS model, STR case and the model we considered in this paper. At higher temperature, entropy of photon gas in the MS model is lesser than special theory of relativity result but bigger than $\kappa$-deformed space-time case. In figure-II, Internal energy is same for the case of $\kappa$-deformed space-time, MS model and normal special theory of relativity case at lower temperature but at higher temperature, internal energy in the MS case is lesser than special theory of relativity result and bigger than $\kappa$-deformed space-time result. However, in figure-III, the specific heat is constant in the special theory of relativity case as temperature increases. At lower temperature, specific heat in MS case and STR case is same and bigger than $\kappa$-deformed space-time model but at higher temperature, specific heat in $\kappa$-deformed space-time case and the MS case is always lesser than STR case. In the MS result, specific heat asymptotically decreases to zero about temperature $T=2000$ but in the our considered model, specific heat asymptotically decreases and goes to zero till about $T=10000$. Hence here we noticed that thermodynamic properties of the photon gas depends on the modification of energy-momentum dispersion relation.

\begin{center}
\includegraphics[height=2.5in, width=4in]{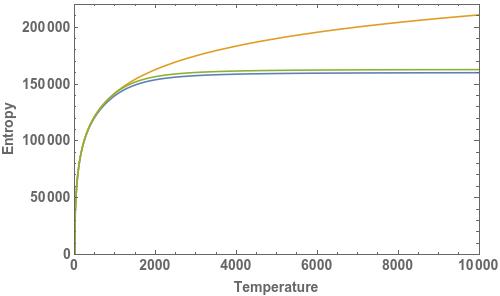}
\end{center}
\textbf{Figure-I:} Entropy S versus Temperature T for three cases: Special theory of Relativity(STR), MS approach and $\kappa$-deformed space-time. Yellow line represents STR, green line) shows the MS result and blue line represents $\kappa$-deformed model.
\begin{center}
\includegraphics[height=2.5in, width=4in]{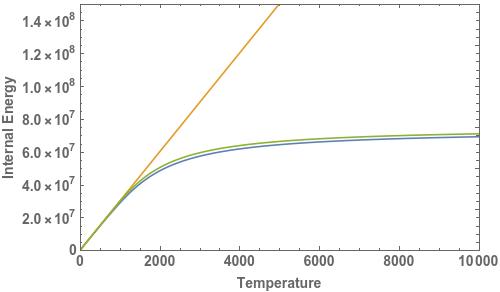}
\end{center}
\textbf{Figure-II:} Internal energy versus Temperature T for three cases: Special theory of Relativity(STR), MS model and $\kappa$-deformed space-time. In this figure, yellow line is correspond to the Special theory of Relativity(STR) result, green line for the MS model and blue line is correspond to the $\kappa$-deformed space-time result. 
\begin{center}
\includegraphics[height=2.5in, width=4in]{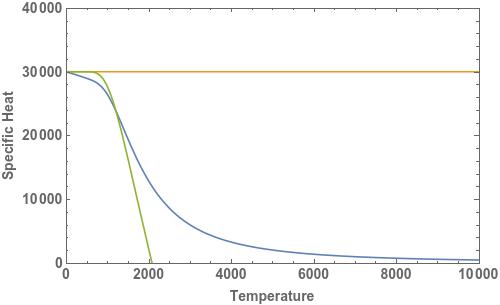}
\end{center}
\textbf{Figure-III:} Specific heat versus temperature for three cases: Special theory of Relativity(STR), MS result and $\kappa$-deformed space-time. In this figure, yellow line is represent the Special theory of Relativity(STR) result, green line is for MS model and blue line is represent the $\kappa$-deformed space-time result.

\section{Discussion}
In this section, we have discussed a brief introduction of work given in\cite{14} and a model we have considered in this paper. DSR leads to study both model MS base dispersion relation as well as $\kappa$-Minkowski space-time.

In \cite{14}, they have started with the MS base dispersion relation given by
\begin{equation}
E^2-p^2=m^2\bigg(1-\frac{E}{\kappa}\bigg)^2.
\end{equation}  
Here $E$ and $p$ are energy and momentum of the particle respectively, $m$ is the mass of the particle and here they have taken $c=1$. Here we have to note that $\kappa$ is the upper bound of the energy scale i.e. Planck energy scale. In the limit $\kappa \rightarrow \infty$, one can get dispersion relation in the commutative case. Also, MS base modified dispersion relation does not change the form for photon and it is same as usual STR scenario. Thus for photons, the dispersion relation becomes
\begin{equation}
E=p.
\end{equation} 
Here note that for the photons form of the dispersion relation is same as STR but upper bound of the enegy scale is $\kappa$ which is missing in the STR scenario. It is clear that in \cite{14}, $p_{max}=\kappa$.

But in this paper, we have considered $\kappa$-deformed dispersion relation upto first order in the deformation parameter and as we discussed in section 3, for photons, dispersion relation is not same as STR scenario but for photon it is same in MS base case. In our considered model $p_{max}$ is different and it is not same as MS case (see section 3 for detail).

Also, we observed  that  entropy of the photon gas has upper bound (i.e. maximum entropy) with respect to temperature variation at high temperature regime which is the purely effect of Non-commutativity. Physically one can be explain as in classical space-time i.e commutative case entropy of photon gas has increasing nature as temperature increases thus this imply us quantum structure of space-time naturally can taming the disorder of the system at high temperature.

\section{Conclusion}
In this paper, we have carried out the thermodynamics of photon gas in the $\kappa$-deformed space-time. For studying this, we started with the $\kappa$-deformed dispersion relation upto first order in the deformation parameter $a$. With this $\kappa$-deformed dispersion relation upto first order in the deformation parameter, we have obtained the partition function of photon gas in the $\kappa$-deformed space-time. This is one of the main findings in this paper. From this $\kappa$-deformed partition function, we next studied the thermodynamic properties of the photon gas in the $\kappa$-deformed space-time and obtained the expression for free energy, equation of state, entropy, internal energy, energy density and specific heat of photon gas in the $\kappa$-deformed space-time. In the limit $\lambda \rightarrow \infty$, we get back the commutative results. Also, we plotted the graphs of entropy, internal energy and specific heat versus temperature for three cases: special theory of relativity, MS model and the $\kappa$-deformed space-time and compared the results. 

As we have studied the thermodynamic properties of photon gas in the $\kappa$-deformed space-time by using $\kappa$-deformed dispersion relation(see Eqn.(\ref{D})). Using this modified dispersion relation due to the $\kappa$-deformation of space-time, we can similarly study the thermodynamic property of an ideal gas. Also, we can study behaviour of fermion gas in the $\kappa$-deformed model. In the $\kappa$-deformed space-time, there might be some modifications in the Fermi energy level which can modify the Chandrasekhar mass limit for the white dwarf stars in the context of noncommutativity\cite{zzz,sayan}. Work along these lines is in progress and shall be reported separately.

\begin{flushleft}
\textbf{Acknowledgements:}
\end{flushleft}
We would like to thanks to referee for their useful comments to make this manuscript more effective.

\end{document}